\documentclass[twocolumn,showpacs,preprintnumbers,amsmath,amssymb]{revtex4}
\usepackage{gensymb}
\usepackage{graphicx}
\usepackage{dcolumn}
\usepackage{bm}
\usepackage[usenames]{color}

\begin{document}


\title{Theoretical investigation of magnetoelectric effects in Ba$_{2}$CoGe$_{2}$O$_{7}$}

\author{Kunihiko Yamauchi$^{1, 2}$}

\author{Paolo Barone$^1$}
\author{Silvia Picozzi$^1$}%
\affiliation{
1. Consiglio Nazionale delle Ricerche (CNR-SPIN), 67100 L'Aquila, Italy \\
2. ISIR-SANKEN, 
 Osaka University, 8-1 Mihogaoka, Ibaraki, Osaka, 567-0047, Japan}


\date{\today}
\newcommand{\ba}{Ba$_{2}$CoGe$_{2}$O$_{7}$}
\newcommand{\bas}{BCGO}
\newcommand{\beq}{\begin{eqnarray}}
\newcommand{\eeq}{\end{eqnarray}}
\begin{abstract}


A joint theoretical approach, combining macroscopic symmetry analysis with microscopic methods (density functional theory and model cluster Hamiltonian), is employed to shed light on magnetoelectricity in \ba. We show that the recently reported experimental trend of polarization guided by magnetic field (H. Murakami et al., Phys. Rev. Lett. {\bf 105}, 137202 (2010)) can be predicted on the basis of phenomenological Landau theory. From the microscopic side, \ba $\:$ emerges as a prototype of a class of magnetoelectrics, where the cross coupling between magnetic and dipolar degrees of freedom needs, as main ingredients, the on-site spin-orbit coupling and the spin-dependent O $p$ - Co $d$ hybridization, along with structural constraints 
 related to the non-centrosymmetric structural symmetry and the peculiar configuration of CoO$_{4}$ tetrahedrons.

\end{abstract}

\pacs{Valid PACS appear here}
\maketitle

The magnetoelectric (ME) coupling between the magnetic order parameter ($M$) and the ferroelectric one ($P$) 
has been well studied in recent years in the context of {\em multiferroic} oxides, where two or more 
primary ferroic phases coexist in the same system.\cite{fiebig.ME.review} 
Most of the research focused onto the {\it linear} ME effect, as expressed by the $\alpha_{ij}H_{i}E_{j}$
cross-coupling term, and observed, e.g.,  in the prototypical
Cr$_{2}$O$_{3}$.\cite{iniguez.cr2o3} 
Microscopically, the ME effect can be ascribed to 
i) the inverse Dzyaloshinskii-Moriya mechanism \cite{sergienko.prb2006} (or the equivalent spin-current
mechanism \cite{KNB}) 
showing $\bm{P}\propto\sum_{ij}\bm{e}_{ij}\times(\bm{S}_{i}\times\bm{S}_{j})$ between neighboring spins
connected by a vector $\bm{e}_{ij}$, and
ii) the inverse Goodenough-Kanamori (or exchange-striction) mechanism \cite{arima.prl2006} showing
$\bm{P}\propto\sum_{ij}J_{ij}(\bm{S}_{i}\cdot\bm{S}_{j})$ with exchange integral $J_{ij}$. 
A third mechanism has been recently proposed, namely 
iii) the spin-dependent $p$-$d$ hybridization\cite{arima.jpsj2007, jia.nagaosa.prb2007}, where the spin-orbit-coupling (SOC)
``asymmetrizes'' the $p$-$d$ hybridization between the transition metal (TM) and the surrounding ligands, inducing an electric
polarization  $\bm{P}\propto\sum_{ij}(\bm{S}_{i}\,\cdot\,\bm e_{j}')^{2}\,\bm e_{j}'$, where $\bm e_{j}'$ labels the
vectors connecting the TM to the ligand ions. 
Since the third mechanism can concomitantly occur with the first or second one (in non-collinear spin structures),
it is in general difficult to identify each of these mechanisms. 
However, it has been recently reported that the third mechanism can be responsible for the polarization observed
in \ba (\bas), where two neighboring Co spins are aligned in an antiferromagnetic (AFM) configuration. 
Incidentally, we have discussed a related ME mechanism in Fe$_{3}$O$_{4}$, 
where 
the crystal structure with $Cc$ space group   doesn't have inversion symmetry, but the $c$-glide symmetry prohibits
polarization along the $y$ direction.\cite{yamauchi.magmagmag} Taking into account SOC and ferrimagnetic spin order, i.e. considering the
magnetic group, the symmetry is broken and a small spin-dependent contribution to polarization arises. 
In this letter, we apply a similar theoretical analysis to \bas: we first perform a symmetry analysis,
then show the DFT results on the ME effect, and further confirm the microscopic mechanims in a model Hamiltonian
approach.

\begin{figure}[ht]
\vspace{-0.2cm}
{
\includegraphics[width=70mm, angle=0]{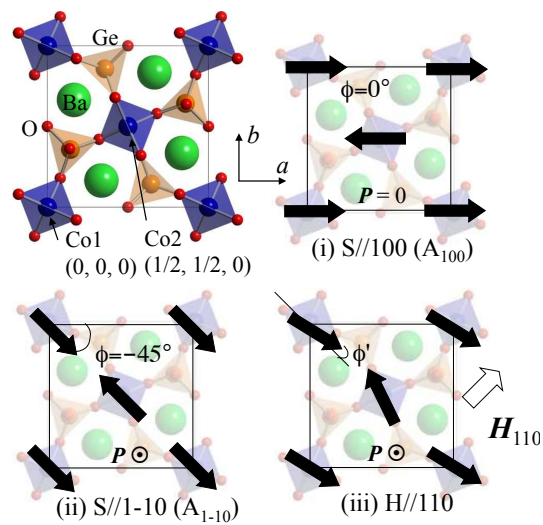}
}
\caption{\label{fig:crys} 
Crystal structure in the ab plane; Co and Ge ions (both located in O$_{4}$ tetrahedron) lie in $c$=0 planes whereas
Ba ion lies in $c$=1/2 planes. 
Co spin configurations: (i), (ii) collinear AFM and (iii) non-collinear spin-canted under
applied $\bm H$//110. 
}
\vspace{-0.3cm}
\end{figure}

\ba\ (melilite) shows tetragonal non-centrosymmetric (but non-polar) $P\overline{4}2_{1}m$ (\#113) structure with
two Co sites, Co1 at (0,0,0) and Co2 at (1/2,1/2,1/2) sites,  as shown in Fig. \ref{fig:crys}. 
Below $T_{N}$=6.7K, the magnetic structure shows collinear AFM spins lying in the $ab$ plane. 
Experimentally, the electric polarization along the $c$ axis, $P_{c}$, is measured even when $\bm H$=0.\cite{cheong.apl2008, tokura.prl2010} 
Additionally, $\bm{P}$ develops finite components along any direction when $\bm H$ is applied,
modulated by the direction and the size of the magnetic field. 

{\em Symmetry Analysis.} --- 
In order to characterize the peculiar ME effect, we first briefly discuss the group theory analysis 
 and its implications in the framework of Landau theory of phase transitions.\cite{Landau} 
In the parent $P\overline{4}2_{1}m1'$ magnetic space group with eight symmetry operations
\{$E$, $C_{2(z)}$, 2$S_{4}$, 2$C_{2(x,y)}$, 2$\sigma_{d}$\}
plus time-reversal (1'), 
the magnetic order leads to a lowered symmetry. 
We define the order parameters  
${\bm F}={\bm S_{1}}+{\bm S_{2}}$ and  ${\bm A}={\bm S_{1}}-{\bm S_{2}}$ as the ferromagnetic (FM) and AFM
combination of Co1 and Co2 spins, respectively. 
\begin{table}[tdp]
\vspace{-0.2cm}
\caption{
Matrices of the generators of space group  $P\overline{4}2_{1}m1'$  in the representations spanned by $F$, $A$ and $P$. 
The group elements denote the identity, $\pi$-rotation, $\pi/2$-rotoinversion, screw $C_{2y}$+$(\frac{1}{2}
\frac{1}{2} 0)$ and time-reversal. Labels of irreducible representation (IR) are taken from the ISODISTORT program.\cite{isodistort}  
\label{table:group}}
\vspace{-0.3cm}
\begin{center}
{\setlength{\tabcolsep}{2pt}\footnotesize
\begin{tabular}{|c|ccccc|c|}
\hline
	& 	$E$  &$C_{2z}$ & $S_{4}^{-}$ & $C_{2y}$&$1'$ &IR  \\
     \hline
%
$\begin{array}{c}F_{a} \\F_{b}\end{array}$ &
$\begin{bmatrix}1 & 0 \\0 & 1\end{bmatrix}$ &
$\begin{bmatrix}-1 & 0 \\0 & -1\end{bmatrix}$ &
$\begin{bmatrix}0 & 1 \\-1 & 0\end{bmatrix}$ &
$\begin{bmatrix}-1 & 0 \\0 & 1\end{bmatrix}$ &
$\begin{bmatrix}-1 & 0 \\0 & -1\end{bmatrix}$ &
$\begin{array}{c}m\Gamma_{5}E_{1}^{*}a \\ m\Gamma_{5}E_{1}^{*}b\end{array}$ \\
$F_{c}$&1&1&1&-1&-1	&$m\Gamma_{4}A$ \\
\hline
%
$\begin{array}{c}A_{a} \\A_{b}\end{array}$ &
$\begin{bmatrix}1 & 0 \\0 & 1\end{bmatrix}$ &
$\begin{bmatrix}-1 & 0 \\0 & -1\end{bmatrix}$ &
$\begin{bmatrix}0 & 1 \\-1 & 0\end{bmatrix}$ &
$\begin{bmatrix}1 & 0 \\0 & -1\end{bmatrix}$ &
$\begin{bmatrix}-1 & 0 \\0 & -1\end{bmatrix}$ &
$\begin{array}{c}m\Gamma_{5}E_{2}^{*}b \\ m\Gamma_{5}E_{2}^{*}a\end{array}$ \\
$A_{c}$&1&1&1&1&-1&$m\Gamma_{1}A$	\\
\hline
%
$\begin{array}{c}P_{a} \\P_{b}\end{array}$ &
$\begin{bmatrix}1 & 0 \\0 & 1\end{bmatrix}$ &
$\begin{bmatrix}-1 & 0 \\0 & -1\end{bmatrix}$ &
$\begin{bmatrix}0 & -1 \\1 & 0\end{bmatrix}$ &
$\begin{bmatrix}-1 & 0 \\0 & 1\end{bmatrix}$ &
$\begin{bmatrix}1 & 0 \\0 & 1\end{bmatrix}$ &
$\begin{array}{c}\Gamma_{5} \\ \Gamma_{5} \end{array}$ \\
$P_{c}$&1&1&-1&-1&1	&$\Gamma_{3}$\\
\hline
\end{tabular}}
\end{center}
\label{default}
\vspace{-0.7cm}
\end{table}
Using the transformation rules given in Table \ref{table:group}, we express the thermodynamic free energies in terms of
all the possible ME 
coupling terms of the form ${\bm P} \cdot {\bm M}^{2}$
which are invariant under symmetry operations:
\begin{eqnarray}
&F_{\rm ME} = c_{\rm A}P_{c}A_{a}A_{b} + c_{\rm F}P_{c}F_{a}F_{b}
\color{black}+ c_{\rm AF}P_{c}(A_{a}F_{a}-A_{b}F_{b})& \notag \\ \color{black}
&+ c_{1}(P_{a}A_{a}F_{c}-P_{b}A_{b}F_{c}) +c_{2}(P_{a}A_{c}F_{a}-P_{b}A_{c}F_{b}), \label{eq:Fm2}& 
%
\end{eqnarray}
while the dielectric energy is  $F_{\rm DE} =- {\bm P}^{2}/2\chi $,
where $c_A, c_{F}, c_1, c_2, c_{AF}$ and $\chi$ (henceforth set as 1) are constants.
$\bm P$ is then evaluated at the minima of $F=F_{\rm ME}+F_{\rm DE}$, reading
\begin{align}\label{eq:P}
P_{a}&= c_{\rm 1}A_{a}F_{c} + c_{\rm 2}A_{c}F_{a},  \qquad
P_{b}=-c_{\rm 1}A_{b}F_{c} - c_{\rm 2}A_{c}F_{b},    \notag \\
P_{c}&=c_{\rm A}A_{a}A_{b} + c_{\rm F}F_{a}F_{b} + c_{\rm AF}(A_{a}F_{a}-A_{b}F_{b}).  
\end{align}
Note that only the first term of $P_{c}$ originates purely from the AFM order, explaining the observed spontaneous
$\bm{P}$, 
whereas other components are allowed only in the presence of the FM order parameter. 

Hereafter we focus on the $P_{c}$ behavior
assuming a canted AFM configuration under an applied magnetic field, i.e., we first simultaneously counter-clock-wise rotate  two
antiparallel Co spins in the $ab$ plane with the angle $\phi$ from the $a$ axis, then we cant spins by an angle
$\phi'$, as depicted in Fig. \ref{fig:crys}. 
Accordingly, we set
${\bm S_{1}}=S\,(\,\cos(\phi+\phi'),\, \sin(\,\phi+\phi'),\, 0)$ and ${\bm S_{2}}=\,S(\,-\cos(\phi-\phi'),\,
-\sin(\phi-\phi'),\, 0)$, ending up with 
\begin{align}
P_{c}(\phi)&=2S^{2}\sin2\phi(c_{\rm A}\cos^{2}\phi'-c_{\rm F}\sin^{2}\phi' -c_{\rm AF}\sin2\phi')\notag \\
       &=2 \alpha S^{2}\sin2\phi\cos(2\phi'-\beta), \label{eqs:Pangle}
\end{align}
where
$\alpha^2=c_{\rm AF}^{2}+(c_{\rm A}+c_{\rm F})^{2}/2$ and $\tan\beta=-(c_{\rm A}+c_{\rm F})/2c_{\rm AF}$.
By neglecting the canting angle $\phi'$, Eq. (\ref{eqs:Pangle}) perfectly reproduces the experimentally observed
dependence of polarization on the spin angle,
being $P_{c}\propto\sin2\phi$ at $T$=2K and $H$=1T.\cite{tokura.prl2010} 
A spontaneous $P_{c}$ can be therefore induced in the A$_{110}$ (A$_{1-10}$) order but not in the A$_{100}$ (A$_{010}$)
one (cfr. Fig. \ref{fig:crys}). 
Analogously to the case of magnetite\cite{yamauchi.magmagmag}, the non-magnetic group lacks the inversion symmetry, but the
symmetries which prohibit $P_{c}$ (e.g. $C_{2y}$ rotation) are broken by the A$_{110}$ magnetic order. 
Starting from the A$_{110}$ order, further symmetry reduction occurs by applying an external $\bm{H}$. 
Indeed, A$_{110}$ order shows $2_{z}'$ point group, which allows non-zero 
$\alpha_{13}$, $\alpha_{23}$, $\alpha_{31}$, $\alpha_{32}$ linear ME components\cite{mecomponent} in such a way that 
$P_{a}$ and $P_{b}$ can be induced by applying $H_{z}$. 
Finally, Eq. (\ref{eqs:Pangle}) at fixed $\phi$ gives the simple $\phi'$-dependence
$P_{c}(\phi')\propto \alpha \cos(2\phi'-\beta),$
where the phase shift depends on the non-zero $c_{\rm AF}$ coefficient. 


{\em DFT analysis} ---
In order to quantitatively confirm the ME behavior and to investigate its microscopic mechanism, we performed DFT
calculations using VASP\cite{vasp} with GGA-PBE potential (we checked our results by using also GGA+$U$\cite{ldau}
potential with $U$=3 or 5 eV for  Co-$d$ state). 
Due to the lack of experimental information on structural parameters, 
we considered the Ca$_{2}$CoSi$_{2}$O$_{7}$ structure\cite{hagiya} and optimize it by substituting atoms (Ca
$\leftrightarrow$ Ba, Si $\leftrightarrow$ Ge) without SOC.
The optimized structures shows $a$=$b$=8.28\AA\ and $c$=5.58\AA, and the tilting angle of CoO$_{4}$ tetrahedron given by
$\kappa$=23.9$^{\circ}$, consistent with experimental values, $a$=$b$=8.41\AA\ and $c$=5.54\AA\cite{cheong.apl2008} and $\kappa$=24$^{\circ}$.\cite{tokura.prl2010}

%
%
%
\begin{table}[h!]
\vspace{-0.5cm}
\caption{
Magnetic anisotropy energy (MAE) (meV/Co) obtained by comparing the total energy with different spin directions under SOC
and for different values of $U$ in the GGA+$U$ scheme. Spin and orbital moment ($\mu_{\rm B}$) are also reported for $S$//(100).
In the rightmost column we report the calculated $P_{c}$ ($\mu C/m^{2}$) for S (L)//110 with fixed atomic structure.  \label{table:mae}
}
\begin{center}
\begin{tabular}{|c|ccc|cc|c|}
\hline
	& 	$E$(100) & $E$(110) & $E$(001) & $S$ & $L$ & $P_{c}$ \\
     \hline
bare GGA& 0 & 0.00 & +0.17& 2.53 & 0.17 & 12.7\\
$U$=3eV    &0 & 0.00 & +0.16  & 2.61 & 0.17 & 12.2 \\
$U$=5eV	&0 & -0.31 & +0.65& 2.75 &0.24  & 10.6\\
\hline
\end{tabular}
\end{center}
\vspace{-0.5cm}
\end{table}

In the CoO$_{4}$ tetrahedra, the Co$^{2+}$ ion shows orbital-quenched
$e_{g}^{2\uparrow}$$t_{2g}^{3\uparrow}e_{g}^{2\downarrow}$$t_{2g}^{0\downarrow}$ states, which causes a very small
magnetic anisotropy, as shown in Table \ref{table:mae}.
The observed magnetically easy $ab$ plane and hard $c$ axis are consistent with experimental report
(S//010 from Neutron diffraction\cite{zheludev.neutron}). 
The small MAE in the $ab$ plane explains why the spins easily follow an applied $\bm H$:  
even under a small magnetic field, the spins flop to be perpendicular to $\bm H$ and then cant in order to reduce the
Zeeman energy. 

%
\begin{figure}[!h]
\resizebox{70mm}{!}
{\includegraphics{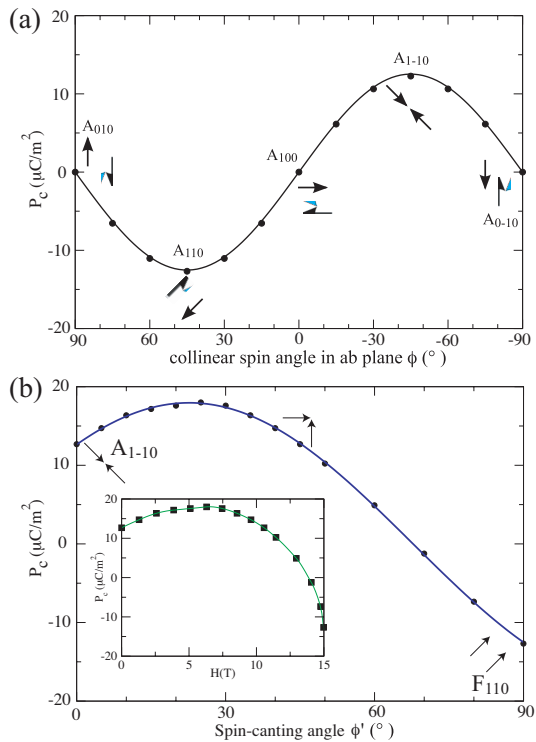}}
\caption{\label{fig:DFT} 
(a) DFT results for 
$\bm P_{c}$ as a function of the {\bf collinear} spin angle $\phi$ in the $ab$ plane,  
fitted to  $f(\phi)=-a\sin2\phi$, with $a$=12.7 (solid line). Spin configurations in the $ab$ plane are shown by arrow. 
(b) DFT results for
$\bm P_{c}$  as a function of the {\bf noncollinear} spin-canting angle $\phi'$ in the $ab$ plane,  
fitted to $f(\phi')=a\cos2(\phi'-b)+c$, with $a$=17.9, $b$=22.7 and $c$=0.06 (solid line). 
Inset: $\bm P_{c}$ as a function of magnetic field. 
}
\vspace{-0.2cm}
\end{figure}
Imposing the collinear AFM configuration, we simultaneously rotate the Co spins in the $ab$ plane. 
We evaluated the ME effect as the change of $\bm P$ (calculated by Berry phase method \cite{berry})
induced by the rotation of $\bm M$ with respect to the crystalline axes, in the fixed non-polar crystal structure. 
In Fig. \ref{fig:DFT} (a) we show $P_{c}$ as a function of the spin-rotation angle $\phi$, consistent with both experiments
and the previously discussed Landau analysis, being $P_{c}\propto\sin 2\phi$. 
The calculated polarization, which displays a maximum value $P_{c}=12.7\,\mu C/m^{2}$ at A$_{1-10}$, originates here from
a purely electronic contribution via SOC, 
and is further enhanced when atomic internal coordinates are optimized in the A$_{1-10}$ configuration, as discussed later. 
%
%

%
%
%
\begin{table}[h!]
\vspace{-0.2cm}
\caption{
$P_{c}$ at different canting angle $\phi'$, calculated in the fixed non-polar structure (first line) and
with optimized internal atomic coordinates (second line). 
Experimentally, the maximum of $P_{c}$ is about 120$\mu$C/m$^{2}$\cite{tokura.prl2010}.  
 \label{table:pcenh}
}
\begin{center}
\begin{tabular}{|c|ccc|cc|c|}
\hline
$P_{c}$	($\mu$C/m$^{2}$)	&$\phi'=$0$^{\circ}$ &30$^{\circ}$ &90$^{\circ}$ \\
     \hline
fixed structure				&12.7	&17.6	&-12.7\\
opt. structure					&39.8	&57.7 	&-38.9	 \\
\hline
\end{tabular}
\end{center}
\vspace{-0.5cm}
\end{table}
We look then at the spin-canting effect induced by an applied field $H_{110}$. 
In Fig. \ref{fig:DFT} (b) we show the change in $P_{c}$ induced by artificially canting the spins by an angle $\phi'$,
starting from the A$_{1-10}$ AFM configuration. 
In agreement with the Landau theory analysis, $P_{c}$ evolves as $\cos2(\phi'-22.7^{\circ})+\rm const.$, 
displaying a peak at $\phi'\sim\kappa$. 
We also evaluate the evolution of $P_{c}$ as a function of the applied $H$ (cfr. inset in Fig. \ref{fig:DFT} (b)),
assuming the experimentally measured magnetic susceptibility $\chi = M/ H_{110}\,\approx\,$0.25 $\mu_{\rm B}$/T per
Co\cite{tokura.prl2010}.
Although the trend of $P$ shows good agreement with experiments\cite{tokura.prl2010}, its size is
one order of magnitude smaller. 
This deviation is reduced when the atomic structure is optimized in the canted-AFM
configuration, as shown in Table \ref{table:pcenh}. 
This means that ferroelectricity is strongly coupled, through magnetism, with lattice distortions 
in a sort of
{\it magnetically-induced piezoelectric effect}. 
The $PH$ curve ---first increasing with $H$ and then decreasing and changing its sign--- denotes an atypical ME trend. 
Although one may assume a non-linear ME coupling coming from high-order terms in the free energy, 
the nontrivial evolution of $P$ actually arises from the composition of the three ${\bm P} \cdot {\bm M}^{2}$ terms
appearing in Eq. (\ref{eq:Fm2}).

{\em Single-site SOC induced ME effect} --- 
The microscopic origin of $\bm{P}$ in \bas~ can be easily explained in terms of a cluster Hamiltonian
for a single CoO$_4$ tetrahedron,
which allows to further clarify the role of the local SOC in the spin-dependent $p$-$d$ hybridization mechanism.
Neglecting contributions from the energetically
deeper majority-spin states, the Hamiltonian consists of four terms, $H=H_d+H_p+H_{pd}+H_{\rm SOC}$, where $H_d
=\Delta \sum_\alpha d^\dagger_\alpha d^{\phantom{\dagger}}_\alpha$ and $H_p =\varepsilon_p \sum_{l,\beta}
p^\dagger_{l,\beta} p^{\phantom{\dagger}}_{l,\beta}$ account for the local energies on Co and O sites (with
$\varepsilon_p=0$ the energy reference and $\Delta=\varepsilon_d-\varepsilon_p$), which
hybridize through
$H_{pd} =\sum_{\alpha,\beta,l}\,V_{\alpha\beta l} (d^\dagger_\alpha\,p^{\phantom{\dagger}}_{l,\beta} + h.c).$
Here $\alpha$ and $\beta$ refer to the $d$ = $xy$, $yz$, $zx$, $x^2$-$y^2$, $3z^2$-$r^2$ and $p= x, y, z$ orbitals involved, whereas
$l=1,...,4$ labels
the four oxygens surrounding the Co ion, located at $\bm{R}_l=(1,1,-1), (-1,-1,-1), (1,-1,1)$
and $(-1,1,1)$ in the local reference system with Co in the origin. The hybridization matrix
$V_{\alpha\beta l}$ depends on the $d$ and $p$ orbitals involved (with $\sigma$ or $\pi$ bonding)
 and on the relative positions of the ions; we adopted the Slater-Koster
parametrization\cite{sk}, assuming 
 $t_{pd\sigma}= 1.3\, eV$,  $\Delta= 5.5\, eV$\cite{vanelp} and $t_{pd\pi}=-0.45\,t_{pd\sigma}$\cite{harrison}. 
\begin{figure}[b]
\vspace{-0.3cm}
\includegraphics[width=7.5cm]{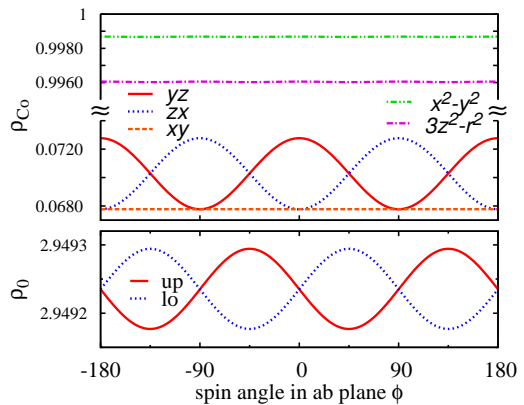}
\vspace{-0.3cm}
\caption{\label{fig:model_charge} Model results. (bottom) Electron density of upper- and lower-lying oxygens,
$\rho_{\mbox{\scriptsize O}}$, in the tetrahedron
model as a function of the azimuthal
angle $\phi$ at $\theta=90 \degree$. (top) Orbital occupancy on Co, $\rho_{\mbox{\scriptsize  Co}}^\alpha$, as a function of $\phi$.}
\vspace{-0.3cm}
\end{figure}
The last term is $H_{\rm SOC}=\lambda\,\sum_{\alpha,\alpha'} \langle \alpha\vert\,L\cdot S\,\vert\alpha'\rangle  \, d^\dagger_\alpha
d^{\phantom{\dagger}}_{\alpha'}$, where the matrix elements can be expressed as a function of the polar and azimuthal
angles ($\theta,\phi$) defining a local reference for the spin-quantization axis\cite{takayama.prb1976}.
We assume $\lambda=0.021 eV$, the free Co ion spin-orbit coupling value. SOC-induced mixing of the local $d$ levels
lifts the degeneracies in the $e_g$ and $t_{2g}$ manifolds and implies different hybridizations
 with the ligand oxygens, that may ultimately induce a local dipole moment. We evaluated
then the local occupancies as $\rho_{\mbox{\scriptsize O}(l)}=\sum_\beta\, \langle\,
p^\dagger_{l,\beta}\, p^{\phantom{\dagger}}_{l,\beta}\,\rangle$ and $\rho_{\mbox{\scriptsize  Co}}=\sum_\alpha\,
\langle\, d^\dagger_{\alpha}\, d^{\phantom{\dagger}}_{\alpha}\,\rangle$ as a function of the azimuthal angle $\phi$, i.e. rotating
the spin in the $ab$ plane.
As shown in Fig. \ref{fig:model_charge}, we can distinguish between
lower-lying (O$_1$, O$_2$) and upper-lying (O$_3$, O$_4$) oxygens, with 
$\rho_{up,lo}\propto \pm\sin(2\phi)$. Then a local dipole
$\bm{p}=(e/4)\,\sum_{l}\, \rho_{\mbox{\scriptsize O}(l)}\, \bm{R}_{\,l}$
may develop only along $c$,
proportional to the charge difference $\Delta\rho_O=\rho_{up}-\rho_{lo}$,
i.e. $p_c\propto 2\sin2\phi$, in excellent agreement with the predicted functional form 
$\bm{P}\propto\sum_{ij}(\bm{S}_{i}\,\cdot\,\bm e_{j}')^{2}\,\bm e_{j}'$
\cite{arima.jpsj2007,jia.nagaosa.prb2007,tokura.prl2010}.
Furthermore, we can estimate
the $d-$orbital mixing on the Co site by looking at the orbital occupancies, shown in Fig. \ref{fig:model_charge}. Even if the
two occupied states have prevalent $d_{x^2-y^2},d_{3z^2-r^2}$ characters, a small mixing occurs via SOC with (mostly)
$d_{yz}, d_{zx}$ orbitals, being  $\rho_{yz}\propto \cos^2\phi,\,\rho_{zx}\propto\sin^2\phi$, i.e. the most occupied is
the one perpendicular to the spin-quantization axis.
\begin{figure}[hb]
\vspace{-0.1cm}
\resizebox{70mm}{!}
{\includegraphics[angle=90,height=5cm]{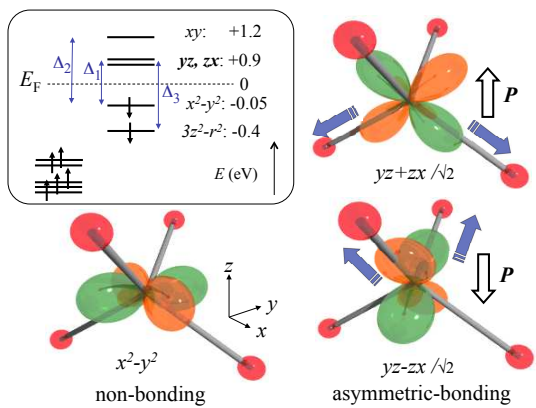}}
\vspace{0.35cm}
\caption{\label{fig:orbital} 
Bonding nature of $d$-orbital in O$_{4}$ tetrahedron and induced local polarization. 
Via SOC, asymmetrically bonded orbital states are mixed with non-bonding occupied states. 
Inset: DFT-calculated energy levels of orbital states. Possible SOC mixing in minority spin states are shown with energy difference $\Delta_{i}$. }
\vspace{-0.4cm}
\end{figure}
\begin{table}[!h] 
\vspace{-0.5cm}
\begin{center}
\caption{DFT-calculated $3d$ orbital-decomposed occupancy (in percentage, with spin states summed up) with different
SOC enhancement factors $\lambda$ (0=without SOC, $\times$1=with standard SOC, and $\times$10=with the  SOC term 10 times
 artificially enhanced) for different $\bm S$ directions in local $xyz$ frame. } 
\label{tbl.coeff}
\begin{tabular}{|cc|ccccc|c|}
\hline 
$\lambda$	&$\bm M$		&$xy$ & ${yz}$ & ${zx}$& $3z^{2}$-$r^{2}$& $x^{2}$-$y^{2}$  \\
\hline
0			&-		&50.0 &50.0 &50.0& 100.0& 100.0  \\
\hline
$\times$1	&$\bm S$//$\bm x$	&50.0 &\bf 50.2& 50.0 & 99.9 & 99.9 \\
$\times$1	&$\bm S$//$\bm y$	&50.0 &50.0& \bf 50.2 & 99.9 & 99.9\\	   
\hline 
$\times$10	&$\bm S$//$\bm x$	&49.5 &\bf 61.7&  49.7& 92.8& 96.3\\
$\times$10	&$\bm S$//$\bm y$	&49.5 & 49.7& \bf 61.7 &92.8& 96.3\\
\hline
\end{tabular}
\end{center}
%
\vspace{-0.4cm}
\end{table}
These findings are in excellent agreement with DFT calculations, as shown in Fig. \ref{fig:orbital} and in Table
\ref{tbl.coeff}, where the hierarchy of $d$-orbital occupancies at selected values of the spin direction is confirmed.
Such a mixing of local $d$-levels nicely explains why $p_c$ size is maximum at $\phi=\pm 45 \degree$,
when the Co spin is parallel either to the upper- or to the lower-lying oxygen bond;  indeed, as
pictorially shown in Fig. \ref{fig:orbital}, the composition of $yz$ and $zx$ orbitals has an asymmetric
bonding nature in the tetrahedron, i. e. non-bonding with upper ligands and bonding with lower ligands or
vice versa.

We can further estimate from our model the $P_c$ dependence on the spin angle in the system by considering
two CoO$_{4}$ tetrahedra tilted by $\kappa$. In the AFM collinear configuration we find $P_c\propto
p_{c1}(\phi+\kappa)+p_{c2}(\phi +\pi - \kappa) = 2\cos\kappa\sin2\phi$.
Analogously, in order to mimick the effect of the external $H_{110}$, we can define the
canting angles as $\phi'_1=\phi+\kappa-\pi/4$ and $\phi'_2=-\phi+\kappa+3\pi/4$,
finding $P_c(\phi')\propto\cos2(\phi'-\kappa)$, in excellent agreement with experiments and DFT results.

{\em Conclusions} --- 

We shed light on the mechanism underlying peculiar magnetoelectric effects in  \ba,
by combining different theoretical approaches and explicitely taking into account the microscopic atomic arrangement and symmetries of the compound. Our Landau phenomenological theory shows that:
i) On top of non-centrosymmetric non-polar  $P\overline{4}2_{1}m$ symmetry in the non-magnetic crystal
structure, a collinear antiferromagnetic spin configuration with in-plane spins allows an electric polarization along the $z$ axis. 
ii) Upon applying an external magnetic field, the induced non-collinear spin-canting well reproduces the experimentally observed peculiar trend of polarization related to the tilting angle between CoO$_4$ tetrahedrons. 
In order to have quantitative estimates, we perform relativistic ab-initio calculations and highlight the delicate interplay between orbital occupation and local magnetic anisotropy, resulting in an excellent match with available experiments. Furthermore, as a proof that the microscopic origin of magnetoelectricity is based on two relevant ingredients (i.e. the anisotropic $p$-$d$ hybridization between Co and O states and the on-site spin-orbit coupling), 
we built a realistic tight-binding model,  
taking into account CoO$_{4}$ tetrahedron and the crystal field splitting, 
that is sufficient to nicely explain magnetoelectric effects and put forward \ba $\:$  as a prototype of the class of materials where the interplay between magnetism and ferroelectricity is based on spin-dependent $p-d$ hybridization, as recently suggested in the literatures 
based on two-ion cluster model.\cite{tokura.prl2010, miyahara.condmat2011} 

During completion of the work, we became aware of a similar symmetry analysis \cite{toledano} performed for BCGO by Toledano {\em et al.}. However, their focus is on toroidal moments, whereas ours is on the combination between single-ion anisotropy and $p-d$ hybridization (derived from density functional and tight-binding models) as main microscopic mechanism driving magnetoelectricity.
 
%

\acknowledgments
Authors thank Y. Tokura and J. Manuel Perez-Mato for fruitful discussions.
The research leading to these results has received funding from the EU Seventh Framework Programme  (FP7/2007-2013) under the ERC grant agreement n. 203523-BISMUTH.
Computational support from Caspur  Supercomputing Center  (Rome) is gratefully acknowledged. 
\end{document}